# Information about Optic Nerve Head Biomechanics is Already Coded in Your Clinical Charts and Can be Used to Predict Glaucoma Progression

Yibo Chen[1,2], Patipol Tiyajamorn[1], Charis Y.N. Chiang[1,3,4], Deepta Ghate[1], Thanadet Chuangsuwanich[1], Michaël J.A. Girard[1,2,5,6,7]

1. Department of Ophthalmology, Emory University School of Medicine, Atlanta, Georgia USA
2. Department of Biomedical Engineering, Georgia Institute of Technology/Emory University, Atlanta, GA, USA
3. Department of Biomedical Engineering, National University of Singapore, Singapore
4. Singapore Eye Research Institute, Singapore National Eye Centre, Singapore
5. Emory Empathetic AI for Health Institute, Emory University, Atlanta, GA, USA
6. Emory Global Diabetes Research Center, Woodruff Health Sciences Center, Emory University, Atlanta, GA, USA
7. Center for Visual and Neurocognitive Rehabilitation, Atlanta VA Medical Center, Atlanta, GA, United States.

**Word count:** 5,259 (manuscript text)

**Tables:** 1

**Figures:** 5
+1 (supplementary material)

**Commercial relationship:** MJAG is the co-founder of the start-up company Abyss Processing Pte Ltd.

**Ethics statement:** This study involves human participants and was approved by the Emory Institutional Review Board.

**Support:** (1) the Emory Eye Center start-up funds, (2) grants from the National Eye Institute, National Institutes of Health (R01EY037299 and R01EY037245), (3) a National Institutes of Health grant to the Atlanta Vision Community (P30EY06360), and (4) a Challenge Grant from Research to Prevent Blindness to the Department of Ophthalmology at Emory University

**Corresponding Author:** Dr Michaël J.A. Girard
Ophthalmic Engineering & Innovation Laboratory (OEIL), Emory Eye Center, Emory School of Medicine, Emory Clinic Building B, 1365B Clifton Road, NE, Atlanta GA 303
mgirard@ophthalmic.engineering
https://www.ophthalmic.engineering

# Abstract

Glaucoma patients routinely undergo optical coherence tomography (OCT) imaging and intraocular pressure (IOP) assessment as part of standard care. Because IOP typically differs across visits, it is possible to map visit-to-visit ONH strains by registering OCT volumes acquired under different IOPs. These strains reflect ONH elasticity and, over longer intervals, also capture growth and remodeling of the ONH. In this study, we evaluated whether such visit-to-visit ONH strain signatures could serve as predictors of glaucoma progression. We retrospectively analyzed 525 eyes from 430 glaucoma patients (238 fast and 287 slow progressors), defining fast progression by a visual field index slope of -1.5% per year or worse. 3D strain maps were derived from baseline and follow-up OCT scans using AI. Predictions using T=1 to T=4 follow-up visits were evaluated by five-fold cross-validation in a long-term cohort (T=1: 1.30 years; T=4: 5.26 years) and short-term testing (T=1: 0.46 years; T=4: 1.25 years). Overall, IOP differed between paired visits in 92% of cases. Long-term T=1 prediction achieved an area under the receiver operating characteristic curve (AUC) of 0.74 and accuracy of 0.72. A single short-term follow-up (T=1) yielded an AUC of 0.75 and accuracy of 0.71; four follow-ups yielded an AUC of 0.87 and accuracy of 0.82. SHapley Additive exPlanations (SHAP) identified neuroretinal rim compression and prelaminar elongation associated with fast progression, with substantial inter-eye variability. Overall, visit-to-visit ONH strains derived from routine OCT predicted glaucoma progression after a single follow-up visit at a mean interval of 0.46 years. These findings suggest that short-term OCT/IOP assessments could provide valuable insights into ONH biomechanics, identify eyes at risk of rapid progression, and reveal distinct strain signatures associated with disease progression.

# Introduction

Glaucoma is the world's leading cause of irreversible blindness characterized by progressive retinal ganglion cell axonal damage [1-3]. In clinical practice, extent of glaucomatous damage is commonly assessed using a combination of structural and functional measures. Structurally, retinal nerve fiber layer (RNFL) thickness, measured from optical coherence tomography (OCT) images, represents a widely used metric [4 5]. Functionally, damage is typically evaluated by visual field loss, with the visual field index (VFI) being one commonly used summary measure. These assessments are repeated over multiple patient visits to estimate the rate of disease progression and thus, personalize treatment to each individual [6-8].

While changes in RNFL thickness and VFI have been used to estimate glaucoma progression with some success, their predictive utility remains limited. Both measures are subject to substantial measurement variability and typically require multiple visits over extended follow-up periods to reliably estimate progression rates, thereby delaying timely identification of patients at high risk of rapid progression [9-12]. Moreover, RNFL thickness exhibits a floor effect in later stages of glaucoma, reducing its sensitivity to detect ongoing progression [13 14]. Because treatment decisions are typically differentiated by whether a patient is determined to be a fast or slow progressor, there is a critical need for reliable methods to identify fast progressors early. Progression status directly informs the aggressiveness of intraocular pressure (IOP)–lowering strategies, including the selection of target IOP and decisions regarding escalation from medical therapy (topical IOP-lowering medications) to laser or surgical intervention [15]. Delays in identifying patients at high risk of rapid progression can therefore hinder timely intervention to prevent irreversible vision loss.

Recently, our group, along with others, has identified optic nerve head (ONH) strain (i.e., deformation) induced by IOP as a potential diagnostic and prognostic biomarker for glaucoma [10 16 17]. This concept is biologically plausible, as mechanical strain can influence axonal integrity and contribute to axonal injury [18-22]. However, direct in vivo mapping of ONH strain typically requires controlled biomechanical stress testing [10 16 22], which is not yet ready for widespread clinical adoption. In clinical practice, glaucoma patients routinely undergo OCT imaging together with IOP measurements. Because IOP naturally fluctuates across visits [23-26], OCT volumes acquired longitudinally capture the ONH under varying mechanical loads. By registering OCT volumes obtained at different time points, it is therefore possible to estimate visit-to-visit ONH strains without the need for an explicit stress test. Over short intervals, these strain estimates may primarily reflect the mechanical response of the ONH to differences in IOP, whereas over longer intervals they may increasingly incorporate tissue growth, axonal atrophy, and remodeling (Fig. 1a). As such, visit-to-visit ONH strain signatures may represent a clinically feasible approach for assessing biomechanical susceptibility and predicting glaucoma progression.

In this study we introduce a novel artificial intelligence (AI) technique to map visit-to-visit ONH strains driven by IOP and/or growth & remodeling. We aim to assess whether these measurements provide independent and clinically meaningful information for predicting glaucoma progression and to identify the strain signatures that characterize fast progressors.

# Methods

This study aimed to determine whether ONH strain, derived from routine clinical imaging, could identify eyes likely to show fast glaucoma progression over longitudinal follow-up. We analyzed a large retrospective dataset of glaucoma patients with longitudinal OCT imaging and clinical records. An AI–based framework was developed to estimate three-dimensional (3D) ONH strain by registering follow-up OCT volumes to baseline volumes. The resulting strain maps were used to train an AI predictive model to classify eyes as fast or slow progressors, and explainable

AI (SHapley Additive exPlanations [SHAP]) analysis was used to identify the tissue-specific strain features contributing to these predictions.

## Data Source

All data were retrospectively accessed from a large clinical OCT imaging database at the Emory Eye Center through CONTINUUM, a secure, HIPAA-compliant cloud platform for clinical imaging research (Institutional Review Board–approved). The dataset spans the period from 2004 to 2024 and includes data collected across multiple hospitals and affiliated clinics within the Emory healthcare system. Subject identifiers were used solely for medical record review to confirm diagnostic and demographic information. Glaucoma diagnoses were further confirmed through chart review by glaucoma specialists. The dataset included OCT imaging, IOP measurements, and visual field test reports collected during routine clinical care.

## OCT Imaging

OCT volumes were acquired using a spectral-domain OCT system (CIRRUS HD-OCT; Carl Zeiss Meditec, Dublin, CA, USA), with each scan centered on the optic nerve head (ONH) and covering a 6 mm × 6 mm field of view. The scan volumes had a transverse resolution of 30 $\mu$m and an axial resolution of 1.95 $\mu$m, yielding voxel dimensions of 200 × 200 × 1024 (transverse × axial).

## Visual Field Testing and Definition of Fast Glaucoma Progression

Visual field testing was performed using an automated perimeter (Humphrey Field Analyzer; Carl Zeiss Meditec, Dublin, CA, USA), which provides VFI. Glaucoma progression was defined based on the slope of the VFI (% per year) calculated using visual field tests acquired after the baseline OCT scan. At least three reliable visual field tests obtained at distinct time points after baseline were required, and tests acquired after the final OCT follow-up visit were also included. Visual field tests were not required to coincide with OCT/IOP visits. The VFI slope was estimated by linear regression of VFI against time. Previous studies have used the VFI slope to distinguish eyes with fast versus slow progression [27-29]. In this study, eyes were classified as fast progressors if the VFI slope was $\leqslant -1.5$ % per year, and as slow progressors if the VFI slope was $> -1.5$% per year.

**Dataset Curation** Glaucoma eyes were identified from the Emory Eye Center database using ICD-10-CM diagnostic codes in the H40 category. Eyes were eligible for inclusion if they had longitudinal valid OCT/IOP visits and visual field test results sufficient to determine glaucoma progression status. A valid OCT/IOP visit required an OCT scan and IOP measurement obtained on the same day. All OCT scans were manually reviewed and excluded if they exhibited: (1) incomplete ONH coverage, (2) large discontinuities or gaps between B-scans, or (3) substantial shadowing artifacts obscuring ONH structures. If multiple scans were available on the same day, the highest-quality scan was selected. Hereafter, “visit” refers to a valid OCT/IOP visit unless otherwise specified. Eyes with other optic neuropathies, such as compressive or ischemic optic neuropathy and optic neuritis, or retinal diseases, such as diabetic retinopathy and macular degeneration, were excluded.

Two overlapping cohorts were defined based on different longitudinal follow-up requirements. Firstly, the long-term cohort was established to evaluate whether longitudinal ONH strain could distinguish fast from slow glaucoma progressors. This cohort included eyes with at least five valid visits, comprising one baseline and at least four follow-ups, without restrictions on follow-up duration. Secondly, the short-term dataset was established to investigate whether strain measured over a short observation period could similarly distinguish fast from slow progressors, potentially enabling earlier identification of eyes at risk of fast-progression. Eyes were included if they had three or more valid visits, comprising one baseline, and at least 2 follow-ups within 1.5 years of baseline. Eyes meeting both follow-up requirements were included in both cohorts, while the full cohort comprised the union of both cohorts, with each eye counted only once.

## AI-Based Strain Mapping

To estimate visit-to-visit ONH deformation and strain from longitudinal OCT scans, we developed an AI-based strain mapping model. For each analysis, the model took only two OCT volumes, a baseline scan and a follow-up scan, and estimated the one-by-one mapping between them, defined as a 3D displacement vector field. The model was optimized independently for each scan pair and did not require prior training on a large dataset. In this sense, the approach used AI with n = 1. The workflow is illustrated in Fig. 1b.

### *Segmentation and BMO Registration*

All OCT volumes were preprocessed using a commercial software platform (REFLECTIVITY; Abyss Processing Pte Ltd, Singapore) [30]. REFLECTIVITY generated 3D segmentations of OCT volumes, identifying major ocular tissues visible in OCT imaging, including eight classes: background, RNFL and prelaminar tissue, ganglion cell layer and inner plexiform layer, other retinal layers, retinal pigment epithelium (RPE), choroid, sclera, and lamina cribrosa. The segmentation results were used to mask out background noise. In addition, segmentations of the RNFL, prelaminar tissue, RPE and choroid-sclera interface (CSI) were used in the design of regularization terms, guiding the alignment of these layers during model training.

In addition to segmentation, REFLECTIVITY performed Bruch's membrane opening (BMO) registration of OCT volumes by applying rigid translation and rotation, such that the BMO center and its associated plane normal were aligned across volumes. The BMO was defined as the endpoint of the posterior RPE boundary on either side of the ONH [31].

### *Deformable Registration and Strain Computation*

Adapted from Implicit Neural Representations for Deformable Image Registration (IDIR) [32], our model automatically registered each follow-up OCT volume to its corresponding baseline scan. IDIR is a coordinate-based neural network approach that represents deformation as a continuous spatial function, mapping 3D coordinates to displacement vectors. Unlike conventional deep learning registration methods, IDIR optimizes the neural network independently for each image pair, without requiring prior training on a large dataset.

Specifically, each spatial coordinate within the baseline ONH volume was provided as input to the network, which predicted its displacement to the corresponding location in the follow-up volume. The intensity at this predicted location was then sampled from the follow-up volume and compared with the baseline intensity. The network was optimized by minimizing an image similarity loss together with physical and geometric regularization terms. The image similarity term encouraged alignment between the warped follow-up and baseline volumes, while the regularization terms constrained local volume changes and promoted alignment of segmented ONH tissue boundaries. The resulting displacement vector field was subsequently used to compute voxel-wise 3D strains. The complete mathematical formulation is provided in the Supplemental Material.

Using the estimated displacement vector field, voxel-wise strain was computed throughout the OCT volume, enabling construction of a longitudinal series of strain maps for each eye. Strain was quantified using three measures: effective strain, first principal strain ($E_1$), and third principal strain ($E_3$). Effective strain provides a scalar measure of local 3D deformation that accounts for both compressive and elongation components with higher values indicating greater mechanical deformations in either direction. The first and third principal strains represent the maximum elongation and compressive deformation, respectively. The morphed volume generated from the displacement vector field was expected to match the baseline volume. The full mathematical formulation of strain computation is provided in the Supplemental Material. 3D visualization of OCT volumes and strain maps was performed using Amira software (Thermo Fisher Scientific, Waltham, MA, USA).

### *Validation of Strain Mapping Accuracy and Robustness*

The proposed strain-mapping framework was validated using synthetic deformation, noise robustness testing, comparison with conventional digital volume correlation (DVC), and scan-rescan repeatability analysis. For synthetic deformation, estimated displacement fields and strain maps were compared with known ground truth, with displacement accuracy quantified using endpoint error. Registration robustness was assessed under speckle noise and block occlusion, while repeatability was evaluated using OCT scans acquired approximately 5 minutes apart. For real clinical OCT data, registration accuracy was assessed using the Dice similarity coefficient between baseline and deformed follow-up tissue segmentations. Detailed validation procedures are provided in the Supplemental Material.

## Visit-to-Visit Strain as a Predictor of Glaucoma Progression

To evaluate whether visit-to-visit ONH strain signatures could predict glaucoma progression, the three strain maps generated for each visit-to-visit pair, including effective strain, $E_1$, and $E_3$, were stacked as multi-channel inputs to a 3D convolutional neural network (CNN). The model was trained to classify eyes as fast or slow progressors (Fig. 1b). Here, T denotes the number of follow-up visits included as model inputs. The prediction model was first constructed using only the first follow-up strain volume ($T = 1$), calculated from the first baseline-to-follow-up scan pair. Additional follow-up visits were then progressively incorporated, ranging from $T = 1$ to $T = 4$, to assess whether increasing longitudinal strain information improved prediction performance over time.

Two performance evaluation strategies were used. First, model performance was assessed using five-fold cross-validation in the long-term cohort. The long-term cohort was randomly partitioned into five folds, with four folds used for training and one fold used for validation in each iteration. This process was repeated five times, such that each fold served once as the validation set. Performance was evaluated using the area under the receiver operating characteristic curve (AUC) and classification accuracy. Accuracy was computed using the optimal Youden threshold, defined as the probability threshold that maximized Youden's index, $\text{sensitivity} + \text{specificity} - 1$, on the validation set.

Second, because larger $T$ values corresponded to substantially longer follow-up durations, We used the short-term cohort to evaluate model generalizability over shorter clinical time scales. For this analysis, models were trained and validated using a five-fold cross-validation strategy. The training and validation dataset comprised eyes from the full cohort excluding those belonging to the short-term cohort. The trained models were then applied to the short-term cohort for evaluation. AUC and classification accuracy were reported for T values ranging from 1 to 4, with classification accuracy computed using the optimal Youden threshold determined from the validation set. The architecture of the prediction network is described in the Supplemental Material.

## Explainable AI to identify what makes a fast progressor

To interpret the prediction model, we aimed to identify the strain signatures that were most important for distinguishing fast from slow progressors. Specifically, we asked whether a particular type of deformation, such as compression or elongation stretch, in a specific ONH tissue or location contributed more strongly to the prediction of fast progression. To address this question, we used SHAP, an explainable AI method that estimates how much each local strain feature contributes to the model prediction [33].

At the individual-eye level, SHAP attribution maps were generated for each strain input channel. These maps identified the local strain features that contributed most strongly to the model prediction for each eye. At each location, we then determined whether elongation strain or compressive strain had the stronger contribution to the prediction. Signed SHAP values were used to determine whether that local strain feature pushed the prediction toward the fast- or slow-progressor class. We also analyzed the local orientation of the dominant strain component, asking, for example, whether important compression occurred perpendicular to the tissue boundary. Because ONH anatomy varies across patients, individual-eye SHAP maps were transformed into a common glaucoma ONH atlas before population-level analysis [17 34]. Here, the atlas can be understood as the most average 3D shape of a glaucoma ONH built from many patient OCT scans. Mapping each eye to this atlas allowed point-to-point matching across patients, so that the same atlas location represented a similar anatomical region in different eyes. For instance, a strain feature located in the temporal-inferior RNFL region of one eye could be compared with the corresponding temporal-inferior RNFL region in another eye. This allowed meaningful population-level averaging of SHAP patterns.

After atlas mapping, SHAP attribution maps were averaged across eyes to identify population-level strain signatures associated with fast progression. The corresponding local strain orientations were also averaged across eyes to identify common anatomical directions of important compression or elongation patterns. Tissue-specific analyses were performed using atlas-based masks for neural tissues and connective tissues. Neural tissues included the RNFL and prelaminar tissue. Connective tissues included the choroid, sclera, and lamina cribrosa. Population-level maps were visualized on temporal-nasal slices, inferior-superior slices, and en face projections.  Additional details of SHAP aggregation, tissue masking, and strain-orientation analysis are provided in the Supplemental Material.

# Results

## Cohort Characteristics

Based on VFI progression rate, 500 eyes classified as fast progressors and 500 eyes classified as slow progressors were randomly selected from our Emory Database. Of these, 525 eyes from 430 patients met all eligibility criteria for inclusion in the full cohort, comprising 238 fast and 287 slow progressor eyes. Demographic characteristics of the full cohort are summarized in Table 1. Overall, the racial composition of the cohort was broadly representative of the diverse clinical population served by the Emory Eye Center in Atlanta.

The long-term cohort comprised 506 eyes (235 fast and 271 slow progressors). The mean follow-up intervals from baseline to the first, second, third, and fourth follow-up visits were 1.30 ± 0.80, 2.57 ± 1.18, 3.93 ± 1.50, and 5.26 ± 1.74 years, respectively.

The short-term cohort comprised 94 eyes (39 fast and 55 slow progressors), each with at least two valid follow-up visits within 1.5 years of baseline. Of these, 38 eyes (13 fast and 25 slow progressors) also had a third follow-up visit within this period, and 20 eyes (8 fast and 12 slow progressors) had a fourth. Accordingly, 94 eyes were evaluated at T = 1 and T = 2, 38 at T = 3, and 20 at T = 4. The corresponding mean intervals from baseline to the first, second, third, and fourth follow-up visits were 0.46 ± 0.35, 0.97 ± 0.38, 0.98 ± 0.28, and 1.25 ± 0.21 years, respectively.

In the full cohort, IOP differed between baseline and follow-up in 92.0% of comparisons, with a mean absolute difference of 3.72 ± 3.55 mmHg. Similar patterns were observed in the long-term and short-term cohorts, with differing IOP values in 92.7% and 89.1% of comparisons and mean absolute differences of 3.71 ± 3.46 and 3.58 ± 3.88 mmHg, respectively.

### Representative Examples of ONH Strain Patterns Across Visits

We demonstrated that ONH strains could be mapped between OCT visits (Fig. 2). In a representative eye with a relatively short follow-up interval (July to December of the same year), IOP increased from 11 to 24 mmHg between visits, accompanied by elevated effective strain primarily localized to the optic cup region (Fig. 2a). These strains may reflect an elastic tissue response to the change in IOP. In a separate eye from another patient with longer follow-up intervals, strain patterns became more widespread across the ONH over time (Fig. 2b). These deformations may reflect progressive tissue atrophy over the longer observation period, although IOP-related deformation may also have contributed, particularly at the third follow-up visit when IOP reached 30 mmHg. The high-strain pattern observed at this visit was reduced at the subsequent visit following a decrease in IOP.

## Prediction of Glaucoma Progression – Slow vs. Fast Progressors

Using longitudinal strain maps as input, the model distinguished fast and slow progressors (Fig. 3a). In long-term cohort cross-validation, the model achieved an AUC of 0.74 ± 0.03 and an accuracy of 0.72 ± 0.02 when only the first follow-up strain map was used ($T = 1,\ time\ interval = 1.30 \pm 0.80\ years$). Performance showed an overall increase in performance as additional follow-up strain maps were included ranging from $T = 1$ to $T = 4$.

Model performance was also evaluated in the short-term cohort (Fig. 3b). Even in this short-term setting, the model achieved an AUC of $0.75 \pm 0.01$ using only the first follow-up strain map ($T = 1, time\ interval = 0.46 \pm 0.35\ years$), suggesting that short-interval OCT visits still captured predictive biomechanical information. Both accuracy and AUC also showed an overall increasing trend with additional follow-up visits ($T = 1$ - $4,\ time\ interval = 0.46 \pm 0.35\ -\ 1.25\ \pm\ 0.21\ years$). Notably, AUC and accuracy were higher in the short-term cohort than in the long-term cohort cross-validation across all follow-up configurations, even reaching 0.87±0.05 at $T = 4$.

## What Makes a Fast Progressor?

For model explainability, spatial SHAP analysis of the short-term (T = 1) model revealed both tissue-specific population-level attribution patterns and individual-eye variability (Fig. 4). At the population level (Fig. 4a), fast progression was associated with distinct neural and connective tissue attribution patterns. In the RNFL, compressive strain oriented approximately perpendicular to the neuroretinal rim contributed strongly toward fast progression across the nasal, temporal, and inferior quadrants, but not the superior quadrant. In the prelaminar region, localized elongation strain oriented predominantly along the inferior–superior axis contributed toward fast progression. In the lamina cribrosa, both elongation and compressive strain contributed toward fast progression, with elongation primarily aligned along the inferior–superior axis and compression oriented approximately normal to the anterior lamina surface. Conversely, strain patterns favoring slow progression were primarily localized to the superior RNFL and superior choroid–sclera region. In the choroid and sclera, elongation was oriented approximately normal to the tissue plane, whereas compression was predominantly transverse.

At the individual-eye level (Fig. 4b), two representative fast progressors exhibited distinct strain attribution patterns. Patient 1 showed predominantly compression-dominant regions (blue), whereas Patient 2 showed predominantly elongation-dominant regions (red). This contrast suggests that compressive strain may be more informative for identifying fast progression in some eyes, while elongation strain may play a greater role in others, highlighting patient-specific variability in biomechanical signatures.

## Validation of Strain Mapping Accuracy

Qualitatively, on real clinical OCT data, registration quality was visualized by overlaying the follow-up scan and the morphed image onto the baseline scan using semi-transparent orange maps. The misalignment of anatomical structures observed before registration was largely eliminated after morphing (Fig. 5a). Quantitatively, on the full cohort, the method achieved a relatively high Dice similarity coefficient (0.80 ± 0.03).

The strain mapping model was also validated using synthetic deformations, accurately recovering both the ground-truth displacement fields and corresponding strain maps (Fig. 5b). Robustness testing showed negligible performance degradation under increasing speckle noise and random block occlusion (Fig. 5c). The proposed method consistently outperformed conventional digital volume correlation (DVC), with substantially lower displacement errors, particularly under complex, multi-axis deformations (Fig. 5d). Finally, scan-rescan repeatability analysis showed minimal strain between repeated OCT scans of the same eye (Fig. 5e).

In addition, the proposed method required only 4.09 ± 1.40 min per scan pair, making it more than six times faster than commercial DVC software, which required over 30 min.

# Discussion

In this study, we developed an AI-based framework to estimate visit-to-visit ONH strain from routine clinical OCT and tested its ability to predict glaucoma progression. ONH biomechanical information could be extracted from paired OCT volumes, and the resulting strain maps alone distinguished fast from slow progressors without additional clinical or demographic inputs. Notably, predictive performance was already strong using only the first follow-up visit (T = 1), with a mean interval of just 0.46 years from baseline, and improved further with additional visits. Spatial SHAP analysis further identified putative population-level signatures of fast progression, such as inferior rim compression, while revealing substantial inter-eye variability in strain patterns.

In our full cohort, IOP differed between visits in approximately 92% of visit pairs, which is expected given the natural fluctuation of IOP over time and potential variation in treatment adherence [22 23 25 26]. These differences provide a natural mechanical perturbation from which ONH deformation can be quantified as non-zero 3D strain between paired OCT scans. Over short intervals, this strain may primarily reflect the acute biomechanical response to IOP change and thus provide a practical measure of ONH robustness, for which no routine clinical test currently exists [10 16 22]. Over longer intervals, however, the measured deformation is more likely to include growth and remodeling in addition to IOP-related effects, capturing 3D structural changes in both neural and connective tissues beyond RNFL thickness alone [18-21].

Importantly, strain maps alone predicted glaucoma progression. Even the first follow-up strain map provided useful discrimination between fast and slow progressors, with both AUC and accuracy above 0.7 in the long-term cohort, and performance improved with additional visits. Because four follow-up visits in the long-term cohort spanned more than five years, when progression may already be detectable by conventional measures, the short-term cohort was particularly informative. Predictive performance was already strong using a single strain map derived from the first follow-up visit (T = 1), obtained on average just 0.46 years after baseline. Performance improved with additional visits, reaching an accuracy of 0.82 and an AUC of 0.87 with four strain maps, all derived from follow-up visits within 1.5 years of baseline. These findings suggest that clustered short-term OCT visits may provide sufficient biomechanical information to estimate progression risk earlier than conventional long-term follow-up [12] and could support a deliberately designed short-term test of ONH robustness under different IOP conditions. For example, visits could be scheduled at different times of day to take advantage of physiologic diurnal IOP variation [35 36]. Alternatively, a monitored short-term withdrawal of topical IOP-lowering therapy, such as a one-week washout, could be considered in prospective studies to generate a larger IOP difference [37 38]. The benefit of additional visits is also biologically plausible, as longer-interval strain may increasingly capture tissue atrophy and remodeling. Our strain-only model performed comparably to prior AI-based progression studies: a recent meta-analysis reported mean AUCs of 0.88 for fundus photography and 0.90 for OCT-based models predicting progression versus no progression [39], while another study distinguishing fast from slow progressors reported a best AUC of 0.83 using multimodal inputs [40]. Future work should test whether adding clinical variables such as IOP history, RNFL thickness, age, and treatment information further improves prediction.

Beyond prediction, an important aspect of our study was the ability to identify and interpret strain features associated with fast glaucoma progression. Using spatial SHAP analysis, a widely used post hoc explainability technique, we identified tissue-specific deformation patterns that contributed to distinguishing fast from slow progressors. When considering only two OCT visits separated by an average of 0.46 years, compression of the peripapillary retinal nerve fiber layer (RNFL) and neuroretinal rim emerged as a prominent feature associated with fast progression, particularly in the inferior, temporal, and nasal quadrants. This finding is biomechanically plausible, as IOP variations can induce compression of ONH neural tissues. Indeed, our previous in vivo studies demonstrated substantial IOP-related deformation, with regional compressive principal-strain magnitudes of approximately -2.1 to -2.8% in combined prelaminar and retinal tissues following acute IOP elevation of approximately 18 mmHg [41], and median prelaminar compressive strain relief of 6.69% following trabeculectomy and an average IOP reduction of approximately 13.5 mmHg [42]. Furthermore, acute IOP elevation of approximately 19.6 mmHg was associated with a reduction in BMO-minimum rim width (BMO-MRW) of 7.51 µm, corresponding to an estimated thickness-direction compressive strain of −4.21% [43]. Alternatively, progressive glaucomatous tissue atrophy may also contribute to the observed compression. However, published rates of RNFL thinning, ranging from approximately 0.36 µm/year in slow structural progressors to 2.87 µm/year in fast structural progressors [44], correspond to estimated thickness-direction compressive strains of −0.22% to −1.77% over 0.46 years. Similarly, reported BMO-MRW thinning of approximately 2.06 µm/year corresponds to an estimated compressive strain of −0.47% over the same period [45]. Although these estimates are not directly equivalent to measured principal strains, and the acute experimental IOP changes were considerably larger than the mean absolute intervisit IOP difference of 3.58 mmHg observed in our short-term cohort, these observations suggest that short-term strain features may capture IOP-dependent biomechanical responses in addition to accumulated structural tissue loss. However, the relative contributions of these mechanisms cannot be distinguished from the present measurements.

Another neural signature associated with fast progression was localized prelaminar elongation. Previous studies reported a maximum elongation strain of 1.55% in prelaminar neural tissue following an average IOP reduction of 11.9 mmHg [46], and predicted a elongation strain of 13.66% following an 8-mmHg IOP increase [47]. In contrast, another study reported a reduction in mean prelaminar tissue thickness from 144.1 to 137.2 µm over 5.4 years, corresponding to an estimated thickness reduction of 0.41% over 0.46 years. Notably, thickness reduction in one direction may coexist with elongation in other directions, and 2.2% of eyes in their study exhibited significant prelaminar tissue thickening [48]. As with RNFL compression, these observations suggest that short-term prelaminar elongation may reflect IOP-dependent biomechanical responses in addition to chronic structural changes. Together, RNFL compression and prelaminar elongation were directionally consistent with axonal strain patterns observed in experimental glaucoma under elevated IOP [49].

In connective tissues, both elongation and compressive strain in the lamina cribrosa contributed toward fast progression. Consistent with these findings, Czerpak et al. reported elongation (+0.83%) and axial/circumferential compressive strains (−0.33%/−0.23%) in the lamina cribrosa following an average IOP elevation of 7.1 mmHg, with several strain compliance measures also associated with visual field loss severity and prior progression rates [50]. Chronic LC remodeling may also contribute, as Girkin et al. reported an average increase in anterior LC depth of 0.82 µm/year over 4.32 years, with greater posterior remodeling associated with faster visual field progression; anterior LC depth was assessed given the limited visibility of the posterior LC boundary on OCT [51]. Overall, our findings suggest that the strain signatures associated with fast glaucoma progression may reflect contributions from both IOP-dependent biomechanical responses and chronic structural remodeling across ONH tissues.

At the population level, these strain signatures highlight putative regions of biomechanical vulnerability that may be relevant to axonal injury, although they represent model-derived attribution patterns rather than direct evidence of tissue damage. Importantly, substantial inter-eye variability was observed, with compression being the dominant feature associated with fast progression in some eyes, whereas elongation was more influential in others. These contrasting patterns suggest that different biomechanical responses may be associated with rapid glaucoma progression across individuals, potentially reflecting heterogeneous mechanisms of axonal injury. If validated, patient-specific strain maps could help identify individualized patterns of biomechanical vulnerability and ultimately support more personalized glaucoma management.

Finally, the proposed strain-mapping framework demonstrated both technical robustness and practical applicability to routine clinical imaging. The model accurately registered longitudinal OCT volumes, achieving a Dice similarity coefficient of 0.80 ± 0.03, and enabled 3D deformation mapping without requiring controlled IOP elevation or dedicated biomechanical testing protocols [10 16 22]. Representative cases further supported the biological plausibility of the estimated strain. In a short-term example, an IOP increase from 11 to 24 mmHg over five months was accompanied by elevated strain in the optic cup (Fig. 3a), consistent with cupping deformation observed in controlled IOP-elevation studies [10 16 21 22 52]. In a separate longitudinal example, high-strain regions expanded and intensified across follow-up visits; although strain in the optic cup decreased after IOP was reduced at the final visit, the overall high-strain region continued to expand (Fig. 3b), suggesting ongoing structural remodeling. Elevated strain also spatially corresponded with RNFL abnormalities, supporting the view that visit-to-visit strain captures both IOP-related deformation and longer-term atrophy-related change. Additional validation using synthetic deformation, noise testing, DVC comparison, and repeatability analysis further supported the reliability of the framework. Because the method requires only paired OCT volumes and tissue segmentation, it could potentially be adapted to other OCT platforms with reliable segmentation.

In this work, several limitations warrant further discussion. First, it remains unknown whether visit-to-visit strain derived from naturally varying IOP is equivalent to the acute biomechanical response measured with a controlled IOP stress test, such as ophthalmo-dynamometry[10 16 22]. Moreover, true ground-truth deformation is unavailable in longitudinal human OCT data, despite validation with synthetic deformation, repeatability testing, and DVC comparison. Second, the current framework measures total deformation between visits and cannot yet separate acute IOP-induced deformation from longer-term growth and remodeling; future work should develop methods to decompose these components and quantify ONH robustness and chronic remodeling separately. Third, this was a retrospective single-site study from Emory, and the findings require validation in larger, independent, and more diverse cohorts, although the use of real-world clinical OCT and IOP data support the feasibility of applying the approach to routine care.

In conclusion, we showed that ONH biomechanical information is already encoded in routine clinical records and can be extracted from OCT scans using AI-based strain mapping. These visit-to-visit strain signatures predicted glaucoma progression, including in a short-term cohort. Our findings suggest that acquiring OCT scans over short intervals and mapping the resulting strain may provide a practical approach for assessing ONH biomechanical robustness and, in turn, help identify eyes at greater risk of rapid progression.

# ACKNOWLEDGEMENTS

Acknowledgement is made to (1) the Emory Eye Center start-up funds, (2) grants from the National Eye Institute, National Institutes of Health (R01EY037299 and R01EY037245), (3) a National Institutes of Health grant to the Atlanta Vision Community (P30EY06360), and (4) a Challenge Grant from Research to Prevent Blindness to the Department of Ophthalmology at Emory University.

**Table 1. Demographic Characteristics of the Study Cohort Compared with the Atlanta Population**

| Characteristic | Study Cohort (N = 430) | Atlanta Population* - Data provided as reference |
|---|---|---|
| **Age, years** | 73.9 ± 12.6 | — |
| **Female sex** | 63.0% (271) | 51.0% |
| **Male sex** | 37.0% (159) | 49.0% |
| **Race** | | |
| African American or Black | 43.7% (188) | 46.0% |
| Caucasian or White | 40.5% (174) | 39.1% |
| Asian | 3.5% (15) | 5.3% |
| Other† | 12.3% (53) | 9.6% |
| **Ethnicity** | | |
| Hispanic or Latino | 2.1% (9) | 6.3% |
| Non-Hispanic or Latino | 77.2% (332) | 93.7% |
| Unknown / Unreported | 20.7% (89) | — |

* Atlanta population estimates obtained from the U.S. Census Bureau QuickFacts (City of Atlanta, Georgia).

† Calculated as the percentage not classified as White, Black, or Asian in the categories reported above.

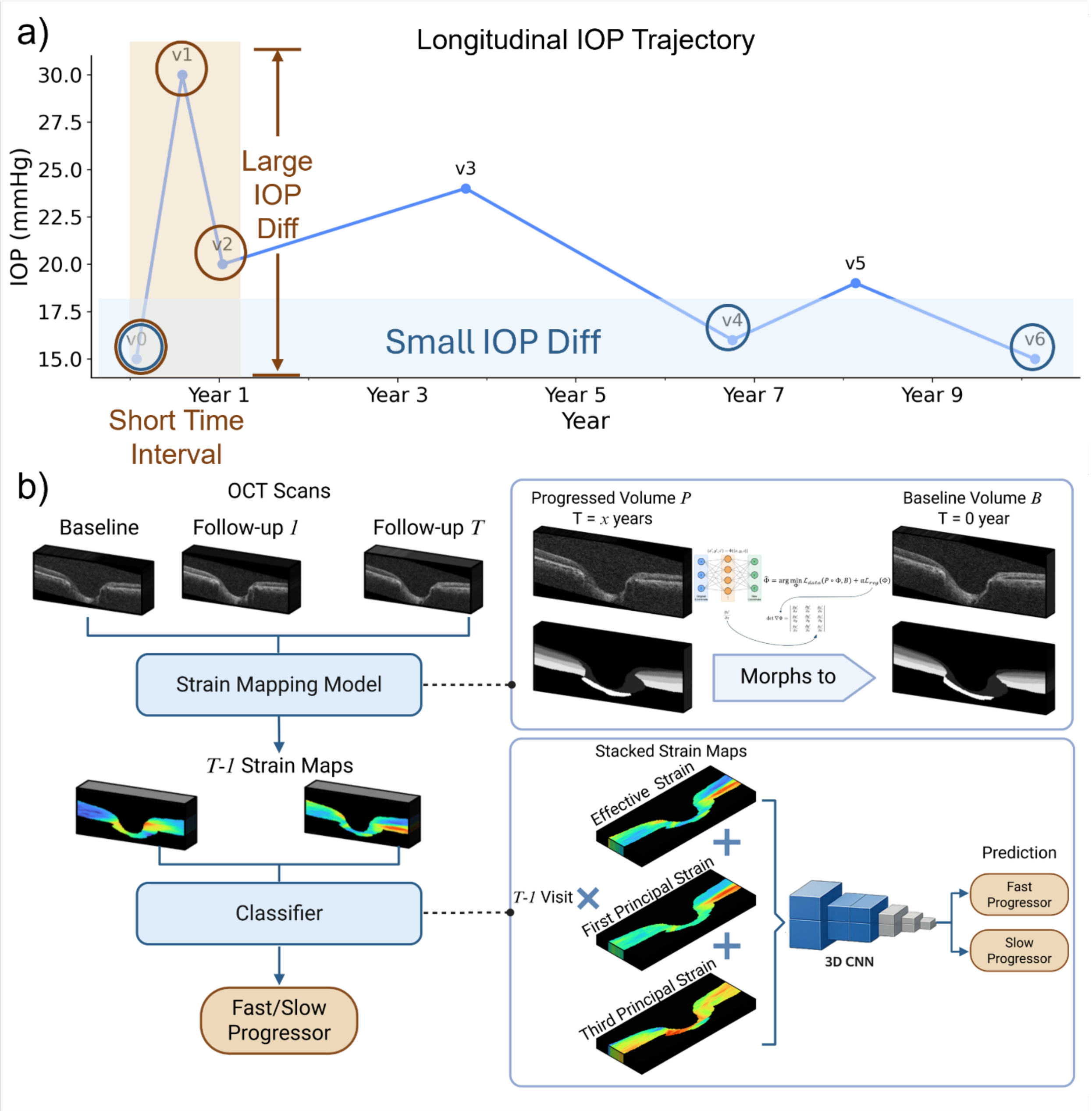

Figure 1. Conceptual motivation and overview of the proposed framework. (a) Schematic longitudinal IOP trajectory illustrating both minor and large IOP differences across clinical visits. These visit-to-visit IOP differences motivated ONH strain estimation from paired OCT volumes. (b) Longitudinal OCT volumes acquired at baseline and follow-up visits are processed using a strain mapping model to estimate deformation between each follow-up scan and the baseline scan. The model outputs displacement vector field s, from which voxel-wise strain maps are computed. For each eye, strain maps from multiple visits are stacked and used as input to a 3D CNN and then the network predicts whether the eye is a fast or slow glaucoma progressor.

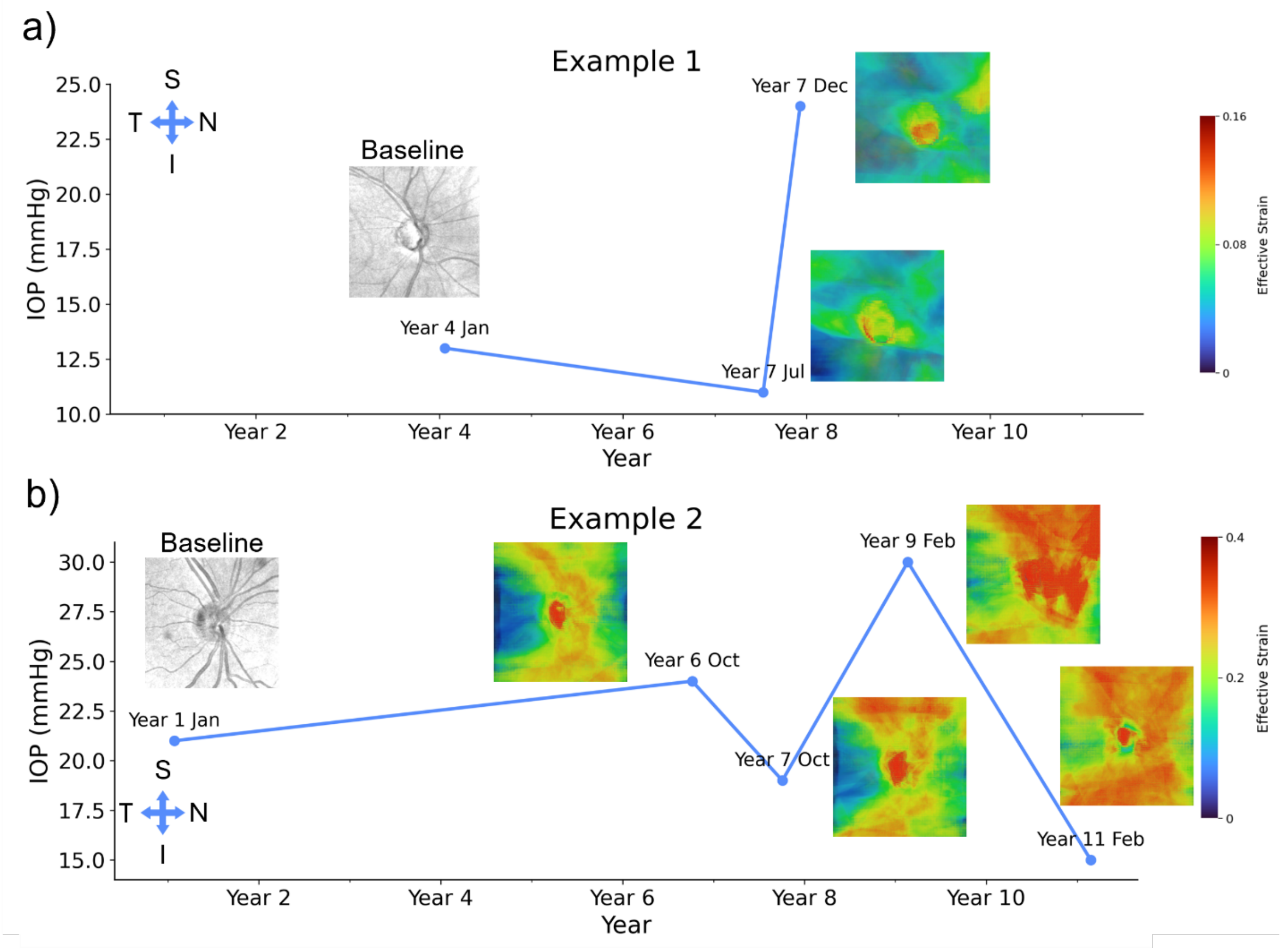

Figure 2. Representative examples of ONH strain mapping across visits. IOP measurements over time are shown alongside baseline OCT images and en face effective strain maps computed relative to baseline. (a) Representative eye with short-term IOP variation. (b) Representative eye from a different patient with longer follow-up intervals and varying IOP. Warmer colors indicate higher effective strain. Color scales differ between panels.

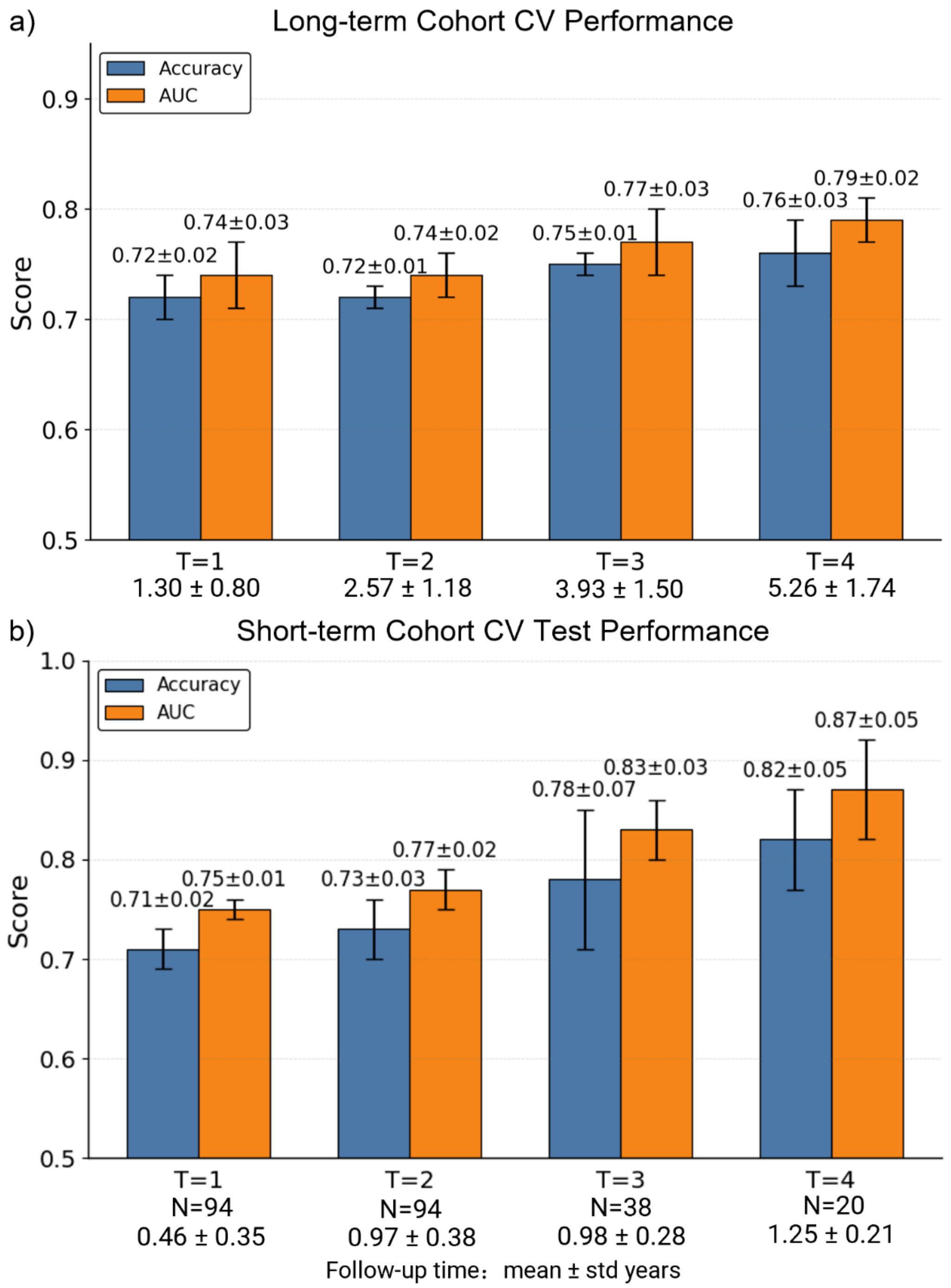

Figure 3. Prediction performance of the proposed model using longitudinal strain maps. (a) Performance on the long-term cohort evaluated using five-fold cross-validation (CV). Accuracy and AUC are shown for models constructed using the first T follow-up strain maps (T = 1–4). (b) Test performance on the short-term cohort (T = 1–4), with accuracy and AUC reported. Error bars represent the standard deviation across the five cross-validation.

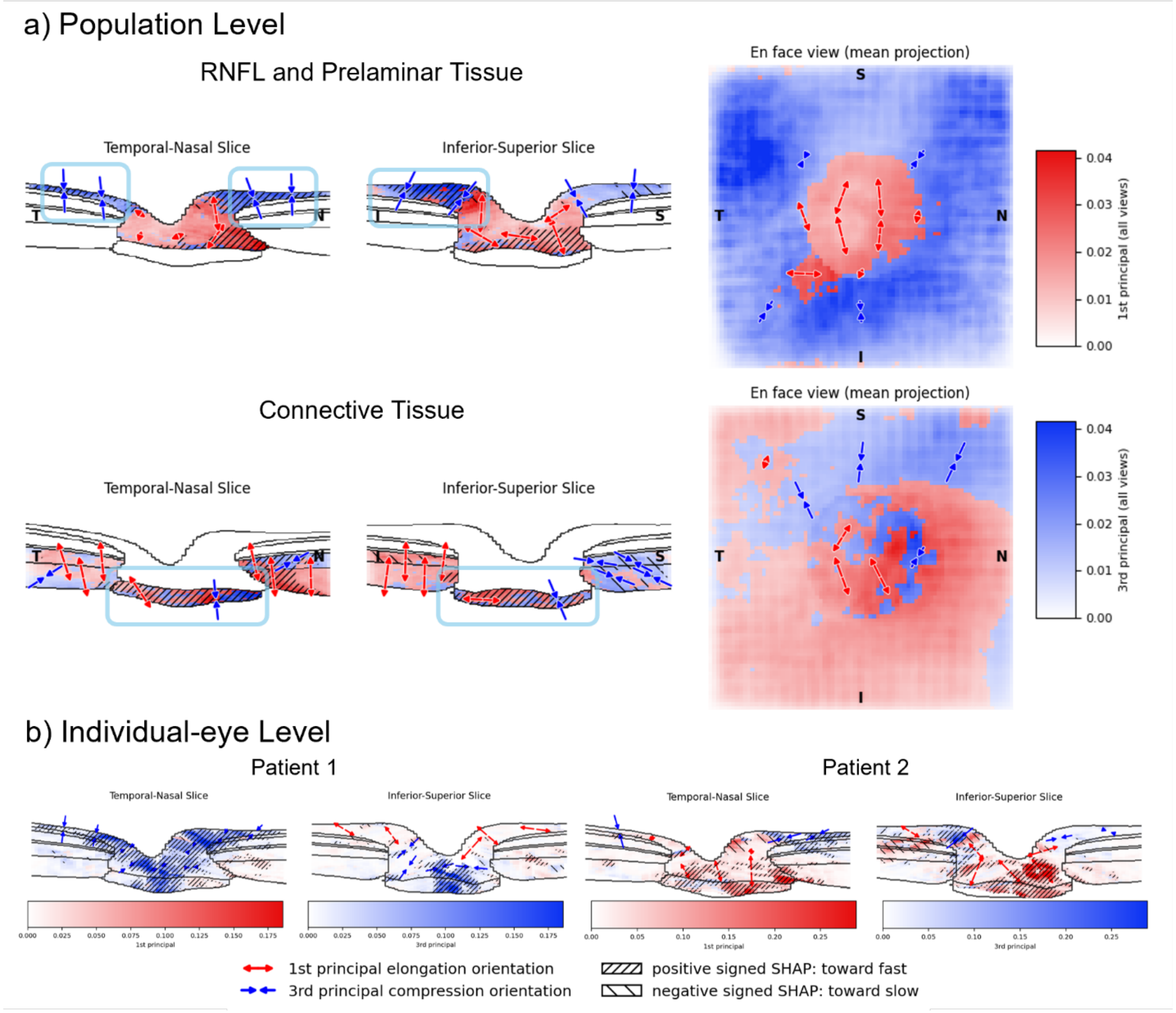

Figure 4. Putative population-level and individual-eye biomechanical signatures associated with fast progression. (a) Population-level SHAP maps after atlas-based averaging across eyes, shown separately for RNFL/prelaminar tissue and connective tissue. Temporal-nasal slices, inferior-superior slices, and en face mean projections are shown. Red regions indicate locations where elongation strain had the dominant attribution, whereas blue regions indicate locations where compressive strain had the dominant attribution. Color intensity represents attribution magnitude. Double-arrow glyphs show local elongation or compression orientation, with arrow length reflecting the in-plane projection of the 3D orientation vector. Hatched regions in the center slices show signed SHAP direction, with positive values contributing toward fast progression and negative values contributing toward slow progression. Light-blue boxes highlight the major regions contributing toward fast progression. (b) Temporal-nasal and inferior-superior slices of representative individual-eye SHAP maps from two patients, showing substantial inter-eye variability in spatial attribution patterns.

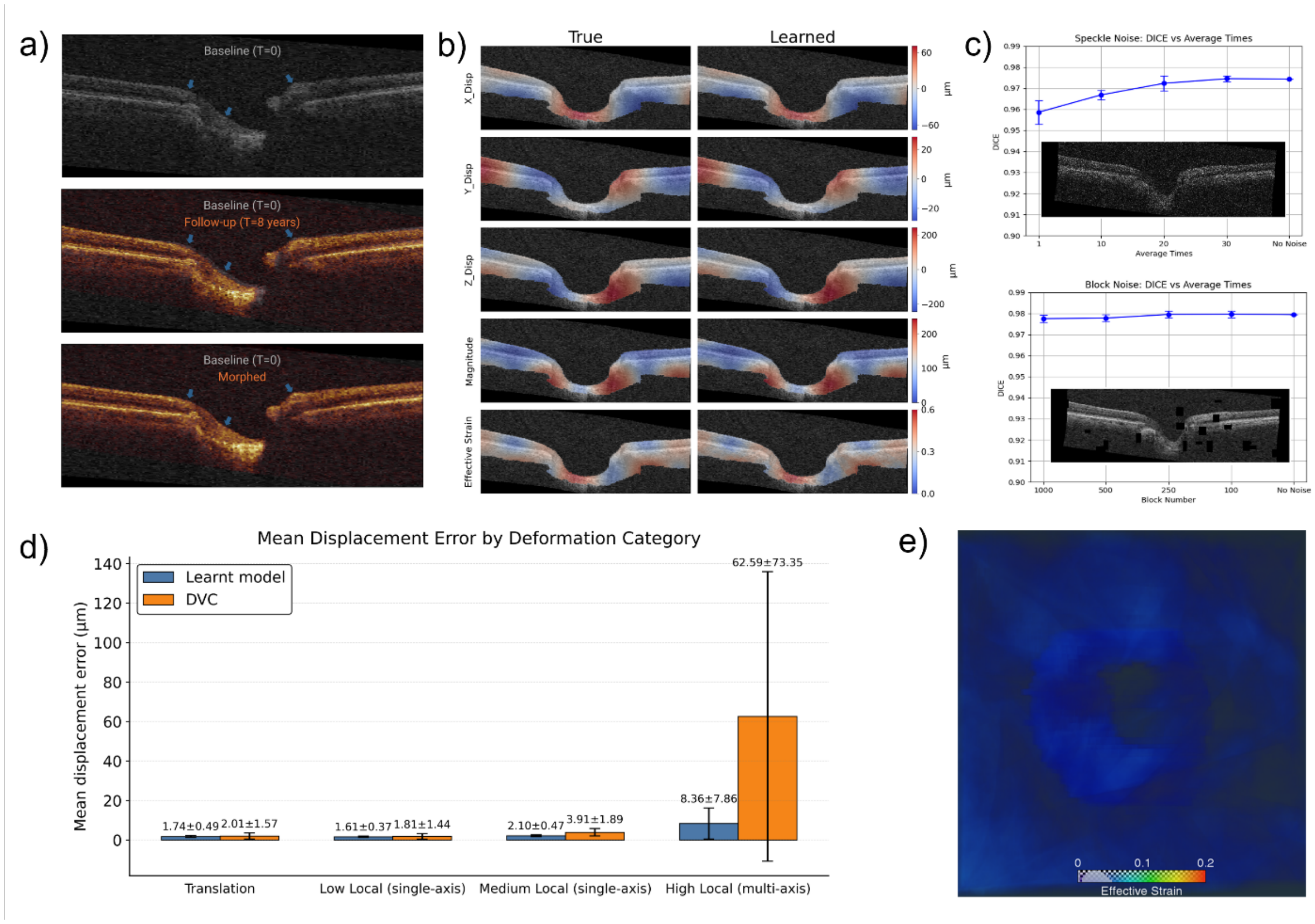

Figure 5. Validation of the proposed strain mapping framework. (a) Qualitative assessment of registration. Follow-up and morphed B-scan images are overlaid onto the baseline scan for the same eye. (b) Validation using simulated deformation. Ground-truth (left) and predicted (right) maps are shown for the displacement vector field components (x, y, z), displacement magnitude, and effective strain. (c) Robustness to noise. Dice coefficient is shown under increasing levels of speckle noise (top) and block occlusion (bottom). (d) Comparison with a conventional DVC method. Mean displacement error is shown across increasing deformation complexity and intensity. (e) Repeatability assessment. Strain map computed from two OCT scans of the same eye acquired within a short interval.

# Supplemental Material

## Deformable Registration and Loss Function

For each baseline and follow-up OCT pair, deformable registration was performed using an implicit neural representation-based framework adapted from Wolterink et al [32]. The baseline volume $B$ was treated as the fixed image, and the follow-up volume $P$ as the moving image. Following the image-sampling convention used in IDIR, the registration network estimated a coordinate transformation $\Phi$ that mapped each baseline coordinate $\mathbf{X}$ to its corresponding sampling location in the follow-up volume:

$$\Phi(\mathbf{X}) = \mathbf{X} + \mathbf{u}(\mathbf{X}),$$

where $\mathbf{u}$ denotes the displacement vector field. The warped follow-up volume, $P \circ \Phi$, was obtained by sampling $P$ at the locations specified by $\Phi$ and was compared with $B$ on the baseline image grid.

The registration network was optimized by minimizing an image similarity loss together with physical and geometric regularization terms:

$$\widehat{\Phi} = \underset{\Phi}{\operatorname{argmin}}\left[\mathcal{L}_{\text{data}}(P \circ \Phi, B) + \mathcal{L}_{\text{reg}}(\Phi)\right].$$

The image similarity loss was defined as the negative zero-normalized cross-correlation (ZNCC):

$$\mathcal{L}_{\text{data}}(P \circ \Phi, B) = -\text{ZNCC}(P \circ \Phi, B).$$

The regularization loss consisted of physical and geometric components:

$$\mathcal{L}_{\text{reg}}(\Phi) = \alpha\mathcal{L}_{\text{phy}}(\Phi) + \beta\mathcal{L}_{\text{geo}}(\Phi),$$

where $\alpha$ and $\beta$ controlled the relative contributions of the two terms.

The physical regularization term constrained local volume changes using the Jacobian determinant of the registration transformation:

$$J_{\Phi}(\mathbf{X}) = \det\left(\nabla_{\mathbf{X}}\Phi(\mathbf{X})\right),$$

$$\mathcal{L}_{\text{phy}}(\Phi) = \frac{1}{N}\sum_{i=1}^{N}|J_{\Phi}(\mathbf{X}_i) - r|,$$

$$r = \frac{V_F}{V_B},$$

where $N$ is the number of sampled points and $V_B$ and $V_F$ denote the combined RNFL and prelaminar tissue volumes in the baseline and follow-up scans, respectively. This term encouraged local volume changes to remain consistent with the observed overall tissue volume change, penalizing excessive local expansion or compression.

The geometric regularization term was based on the Chamfer distance between anatomical point clouds sampled from segmented tissue boundaries [53]. Point clouds were generated from the RNFL, prelaminar tissue, retinal pigment epithelium (RPE), and choroid–sclera interface (CSI) in both baseline and follow-up volumes (Supplemental Fig. 1a). The RNFL and prelaminar tissue were emphasized because of their relevance to glaucomatous axonal damage [4 5 54]. The RPE provided a well-defined retinal boundary to facilitate alignment of large-scale deformations, while the CSI provided an additional posterior connective-tissue boundary.

The geometric regularization term was defined as

$$\mathcal{L}_{\text{geo}}(\Phi) = \mathcal{L}_{B\to F} + \mathcal{L}_{F\to B},$$

$$\mathcal{L}_{B\to F} = \frac{1}{|\mathcal{P}_B|}\sum_{\mathbf{p}\in\mathcal{P}_B}\min_{\mathbf{q}\in\mathcal{P}_F}\| \Phi(\mathbf{p}) - \mathbf{q} \|_2^2,$$

$$\mathcal{L}_{F\to B} = \frac{1}{|\mathcal{P}_F|}\sum_{\mathbf{q}\in\mathcal{P}_F}\min_{\mathbf{p}\in\mathcal{P}_B} \parallel \mathbf{q} - \Phi(\mathbf{p}) \parallel_2^2.$$

where $\mathcal{P}_B$ and $\mathcal{P}_F$ denote the baseline and follow-up point clouds, respectively. The transformation $\Phi$ mapped baseline boundary points to corresponding locations in the follow-up volume. This bidirectional nearest-neighbor loss encouraged alignment of anatomical boundaries and improved geometric consistency (Supplemental Fig. 1b).

## Strain Computation

For biomechanical strain computation, the baseline and follow-up OCT volumes were defined as the reference and current configurations, respectively. The coordinate transformation $\Phi$ estimated during registration corresponds to the forward deformation mapping $\boldsymbol{\varphi}$ in continuum mechanics when both are expressed in the same physical coordinate system. Thus, $\boldsymbol{\varphi} = \Phi$, and the deformation gradient was computed with respect to the reference coordinates.

Following the notation of continuum mechanics, the deformation mapping was expressed as

$$\mathbf{x} = \boldsymbol{\varphi}(\mathbf{X}) = \mathbf{X} + \mathbf{u}(\mathbf{X}),$$

where $\mathbf{X}$ and $\mathbf{x}$ denote the material coordinates in the reference configuration and spatial coordinates in the current configuration, respectively, and $\mathbf{u}$ is the forward displacement vector field.

The deformation gradient tensor was defined as

$$\mathbf{F} = \mathrm{Grad}\boldsymbol{\varphi} = \frac{\partial \mathbf{x}}{\partial \mathbf{X}} = \mathbf{I} + \mathrm{Grad}\mathbf{u},$$

where $\mathbf{I}$ is the second-order identity tensor. The Green–Lagrange strain tensor was then computed as

$$\mathbf{E} = \frac{1}{2}(\mathbf{F}^T\mathbf{F} - \mathbf{I}).$$

Three voxel-wise strain measures were derived from $\mathbf{E}$: effective strain, first principal strain, and third principal strain. Effective strain was defined as

$$E_{\text{eff}} = \sqrt{\frac{2}{3}\,\mathbf{E}:\mathbf{E}},$$

where : denotes the double contraction of two second-order tensors. The first and third principal strains, $E_1$ and $E_3$, were defined as the largest and smallest eigenvalues of $\mathbf{E}$, respectively:

$$E_1 = \max\big(\text{eig}(\mathbf{E})\big),$$

$$E_3 = \min\big(\text{eig}(\mathbf{E})\big),$$

with $E_1 \geq E_2 \geq E_3$. The first and third principal strains characterized the maximum and minimum principal strains, respectively, while effective strain provided a scalar measure of overall strain magnitude. These measures were computed throughout the OCT volume for each baseline-to-follow-up pair, generating longitudinal strain maps for subsequent analysis.

## Artificial Deformation Experiments

Artificially deformed OCT volumes were generated from baseline 3D OCT scans and their corresponding label maps using analytic displacement vector fields. For local deformation experiments, smooth sinusoidal displacement fields were defined along the three spatial directions.

$$u_x = A\sin(k_x x)\cos\big(k_y y\big)\cos(k_z z)$$

$$u_y = B\cos(k_x x)\sin\big(k_y y\big)\cos(k_z z)$$

$$u_z = C\cos(k_x\, x)\cos\left(k_y y\right)\sin(k_z z)$$

Where $(x, y, z)$ are voxel coordinates, $u_x$, $u_y$, and $u_z$ are the displacement components, and $A$, $B$, and $C$ control the displacement amplitude along each axis.

Translation cases were generated using constant displacement fields along one spatial direction. The deformed OCT intensity volumes were generated by applying the analytic displacement field to all voxel coordinates and resampling the baseline OCT volume using trilinear interpolation. The corresponding label maps were resampled using nearest-neighbor interpolation to preserve integer tissue labels. The analytic displacement field was saved as the ground-truth deformation field and was used to compute the ground-truth displacement and strain maps.

For comparison with conventional DVC, experiments were grouped into four deformation categories according to deformation type and intensity: translation, low local deformation, medium local deformation, and high local deformation. Translation cases used constant displacement fields. Low and medium local deformation cases used single-axis sinusoidal deformations with different displacement magnitudes. High local deformation cases used multi-axis sinusoidal deformations to represent more complex deformation patterns. Displacement accuracy was quantified using endpoint error, defined as the mean Euclidean distance between the estimated and ground-truth displacement vectors. For DVC comparison, the same synthetic deformation cases were processed using a global DVC method implemented in Amira, and the mean and standard deviation of displacement errors were evaluated against the same ground truth (Fig. 5d).

## Noise Robustness Experiments

Noise robustness experiments were performed to evaluate the sensitivity of the registration model to image noise and signal dropout. Speckle noise was simulated as multiplicative noise, in which voxel intensities were multiplied by random variables sampled from an exponential distribution. Different speckle levels were generated by averaging multiple independently corrupted volumes, with averaging numbers of 1, 10, 20, and 30, plus a no-noise condition.

Random block occlusion was used to simulate localized signal dropout. Cuboid blocks with randomly selected locations and sizes were inserted into the 3D OCT volume, and voxel intensities within these blocks were set to zero. Block occlusion levels included 1000, 500, 250, and 100 blocks, plus a no-noise condition. For both noise models, noise was applied to the artificial deformed OCT volume before registration, and robustness was evaluated using the Dice similarity coefficient. The Dice coefficient remained high across all noise levels, increasing from 0.959 at the highest speckle noise level to 0.974 under the no-noise condition, while ranging only from 0.978 to 0.979 across different levels of block occlusion.

## Noise Robustness Experiments

To assess registration repeatability, two OCT scans acquired from the same eye approximately 5 minutes apart were registered using the proposed framework. Because no substantial tissue deformation was expected between consecutive scans, the resulting strain was expected to be minimal, primarily reflecting small variations in eye positioning and scanning angle.

## Prediction Network Structure

The prediction network was implemented as a shallow 3D CNN consisting of an initial 3D convolution–batch normalization–ReLU–max pooling block, followed by two additional 3D convolution–batch normalization–ReLU blocks with intermediate 3D max pooling. Adaptive average pooling was applied before a dropout-regularized linear classification head that produced a single logit. The model was trained using weighted binary cross-entropy loss, AdamW optimization, ReduceLROnPlateau learning-rate scheduling, and early stopping based on validation AUC.

## Spatial SHAP Attribution and Principal-Strain Orientation Analysis

Spatial model attribution was quantified using SHAP values. For each eye and each temporal setting $T$, SHAP attribution maps were generated for the effective strain, first principal strain ($E_1$), and third principal strain ($E_3$) channels. Attribution maps were aggregated across cross-validation folds and transformed into a common atlas coordinate system using the inverse displacement field from each eye to the atlas. Population-level attribution maps were then obtained by averaging the atlas-warped attribution maps across all eyes in the short-term cohort.

For tissue-specific visualization, atlas segmentation labels were used to define anatomical masks. Attribution values outside the selected tissue mask were set to zero before identifying the dominant attribution component. At each atlas voxel, the dominant mechanical component was selected by comparing the attribution magnitudes of $E_1$and $E_3$. Effective strain was excluded from this winning-component comparison for visualization because it did not show a dominant attribution pattern in our experiments. The winning component was defined as

$$c^*(x) = \arg \max_{c \in \{E_1, E_3\}} A_c(x),$$

where $A_c(x)$is the nonnegative SHAP attribution magnitude for component $c$ at atlas location $x$. The background color of each voxel encoded the magnitude of the winning attribution component, with red corresponding to $E_1$ and blue corresponding to $E_3$. Center slices and en face projections were scaled separately. Within each multi-row figure, all center slices shared a common intensity scale, and all en face projections shared a common intensity scale.

Signed SHAP values were analyzed separately from the attribution magnitudes used for component selection. After the winning component was identified at each voxel, the signed SHAP value corresponding to that same component was extracted. Positive signed SHAP values indicated that the local feature contribution increased the model score toward the fast-progressor class, whereas negative signed SHAP values indicated contribution toward the slow-progressor class. In the visualization, positive signed SHAP regions were shown with forward-slash hatching, and negative signed SHAP regions were shown with backslash hatching. Hatch overlays were thresholded using a percentile of the absolute signed SHAP distribution to emphasize spatially coherent high-magnitude signed contributions.

En face maps were generated by projecting the masked atlas attribution maps along the anterior-posterior axis, typically using mean projection. For each component, the projected attribution map was computed first, and the winning component in the en face view was then selected by comparing the projected $E_1$ and $E_3$ attribution magnitudes.

Principal-strain orientation was summarized at predefined fixed atlas locations. Separate fixed-point sets were used for neural and connective tissue center-slice analyses, while the same en face fixed-point set was used for both tissue groups. At each fixed atlas point, a local neighborhood was sampled. For center slices, the neighborhood was a local 3D window around the selected point. For en face points, the neighborhood used a local window in the transverse directions and all depths along the projection axis. For each fixed point, the local winning attribution component determined whether the $E_1$ or $E_3$ eigenvector field was sampled. Native-space principal-strain eigenvectors were loaded for each eye. The atlas inverse displacement field was then used to map the fixed atlas neighborhood into each eye's native coordinate system, where the corresponding eigenvectors were sampled.

To compute a population mean orientation, sampled unit vectors were accumulated using an orientation tensor:

$$\mathbf{M} = \sum_{i} v_i \, v_i^{\mathsf{T}},$$

where $v_i$ is a sampled unit eigenvector. The dominant eigenvector of $M$ was used as the mean local orientation. This tensor-based approach treated $v$ and $-v$ as equivalent orientations and avoided cancellation caused by arbitrary eigenvector sign. Fixed-point orientations were plotted as double-arrow glyphs: red outward arrows indicated elongation ($E_1$) orientation, and blue inward arrows indicated compressive ($E_3$) orientation. Because strain orientations were estimated in 3D but visualized on 2D slices or projections, each double-arrow glyph represents the projection of the local 3D orientation vector onto the displayed plane. Therefore, shorter arrows indicate that a larger component of the orientation is perpendicular to the displayed plane, whereas longer arrows indicate that the orientation lies more strongly within the plane.

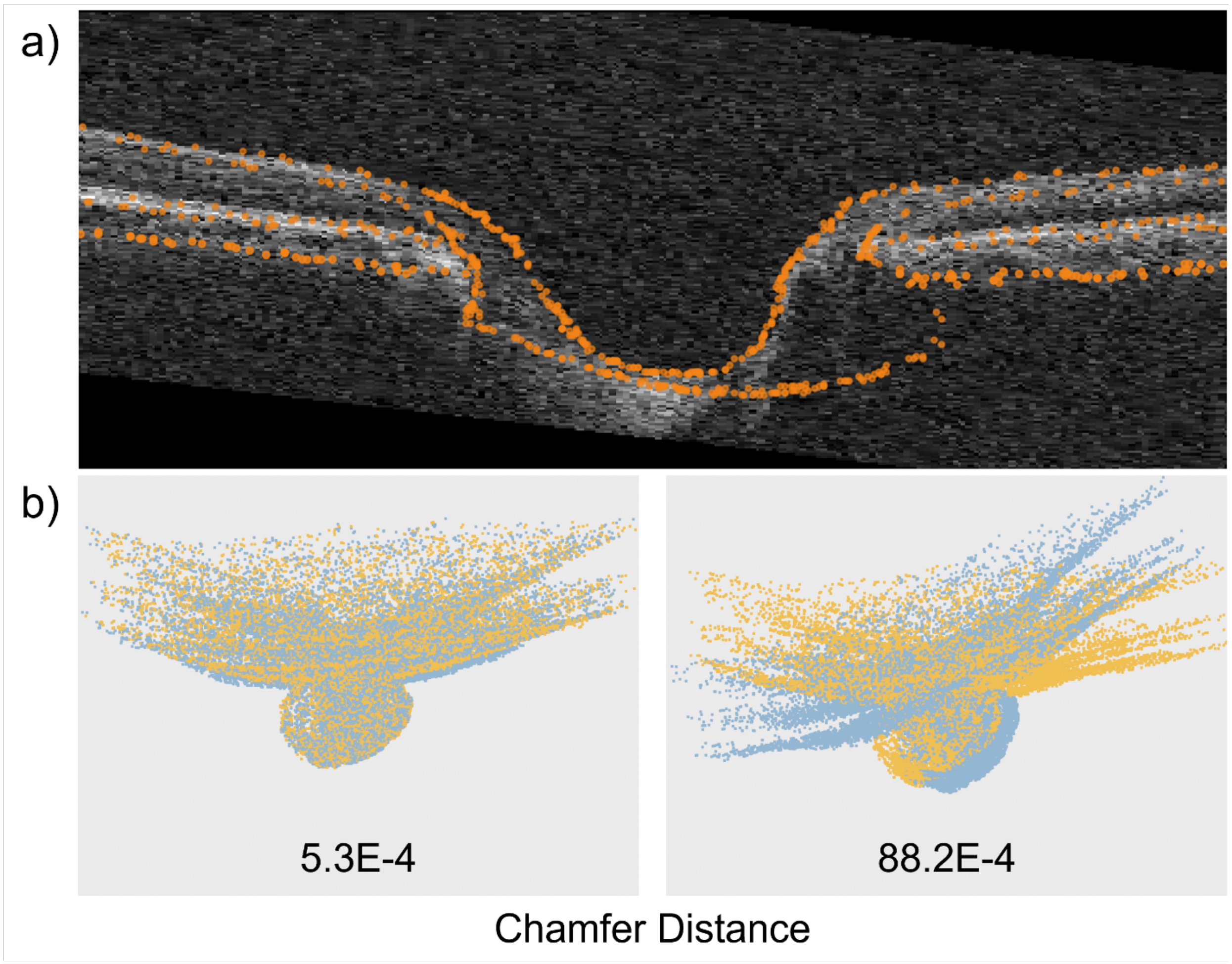

Supplemental Figure 1. Construction of point clouds and boundary alignment using Chamfer distance. (a) Segmentation boundaries of the RNFL, prelaminar tissue, RPE and CSI were sampled to generate point clouds used for geometric regularization during registration. (b) Chamfer distance quantifies the alignment between two point clouds.